\documentclass[twocolumn]{emulateapj}
\usepackage{apjfonts,amsmath}
\usepackage{amsmath}

\newcommand{\Msun}{M_{\odot}}

\begin{document}

\title{Tidal disruption of dark matter halos around proto-globular clusters}

\author{Takayuki R. SAITOH}
\affil{National Astronomical Observatory of Japan, Mitaka, Tokyo
181-8588, Japan\\
E-mail:saitoh.takayuki@nao.ac.jp}
\author{Jin KODA}
\affil{1. National Astronomical Observatory of Japan, Mitaka, Tokyo
181-8588, Japan\\
2. California Institute of Technology, MS105-24, Pasadena, CA91125, USA\\
E-mail:koda@astro.caltech.edu}

\author{Takashi OKAMOTO}
\affil{1. National Astronomical Observatory of Japan, Mitaka, Tokyo
181-8588, Japan\\2. Department of Physics, University of Durham, South Road, Durham DH1 3LE, England\\
E-mail:Takashi.Okamoto@durham.ac.uk}

\author{Keiichi WADA$^1$}
\affil{National Astronomical Observatory of Japan, Mitaka, Tokyo
181-8588, Japan\\
E-mail: wada.keiichi@nao.ac.jp}
\altaffiltext{1}{Department of Astronomical Science, The Graduate University for Advanced Studies, Osawa 2-21-1, Mitaka, Tokyo 181-8588, Japan.}

\and

\author{Asao HABE}
\affil{Division of Physics, Graduate School of Science,
Hokkaido University, N10W8, Sapporo 060-0810, Japan\\
E-mail:habe@astro1.sci.hokudai.ac.jp}

\received{2005 Aug 9}
\accepted{2005 Nov 22}

\begin{abstract}
Tidal disruption of dark matter halos around proto-globular clusters 
in a halo of a small galaxy is studied in the context of the hierarchical 
clustering scenario by using semi-cosmological N-body/SPH simulations assuming 
the standard cold dark matter model ($\Omega_0 = 1$).
Our analysis on formation and evolution of the galaxy and its substructures archives until $z = 2.0$.
In such a high-redshift universe, 
the Einstein-de Sitter universe is still a good approximation 
for a recently favored $\Lambda$-dominated universe,
and then our results does not depend on the choice of cosmology.
In order to resolve small gravitationally-bound clumps around galaxies 
and consider radiative cooling below $T = 10^4 K$, we adopt a fine mass 
resolution ($m_{\rm SPH} = 1.12 \times 10^3 \Msun$).
Because of the cooling, each clump immediately forms a `core-halo' 
structure which consists of a baryonic core and a dark matter halo.
The tidal force from the host galaxy mainly strips the dark matter halo 
from clumps and, as a result, theses clumps get dominated by baryons.
Once a clump is captured by the host halo, its mass drastically decreases 
each pericenter passage.
At $z = 2$, more than half of the clumps become baryon dominated systems 
(baryon mass/total mass $> 0.5$). Our results support the tidal evolution 
scenario of the formation of globular clusters and baryon dominated dwarf 
galaxies in the context of the cold dark matter universe. 
\end{abstract}
\keywords{
galaxies: formation --- galaxies: halos --- method: numerical --- dark matter 
}

%
\section{Introduction}
%

Structure formation in the cold dark matter (CDM) dominated universe 
is characterized by hierarchical clustering,
where small objects collapse early, and merge into larger systems, 
such as galaxies and clusters of galaxies. 
Therefore, the formation process of small-scale structure is crucial 
to understanding the formation of larger objects.
The tidal effect from larger objects could affect star formation history 
in each subhalo (gravitationally-bound systems in galaxies)
and the final structure of the host objects.

$N$-body simulation is a suitable method to follow 
the co-evolution of substructure and host halos.
\citet{kravtsov04} studied the evolution of dark matter subhalos 
within galactic CDM halos using $N$-body simulations without a gas component.
They found that the tidal force is essential for the evolution of 
substructure on galactic scale. 
Baryons, however in general, have more concentrated distribution than dark matter 
owing to their dissipative nature, and hence evolution of these subhalos should be strongly 
influenced by baryonic processes. 

Quantitative behavior of tidal interactions between 
a three-component (stars, gas, and dark matter) substructure 
(hereafter clump) and a host halo can be directly studied by numerical simulations.
However, 
typical mass resolutions in previous
$N$-body/SPH (Smoothed Particle Hydrodynamics) simulations of galaxy formation 
is $10^5-10^7 \Msun$ in mass, $\sim 1$ kpc in length, and 
the number of particles in a halo is typically $10^3-10^5$ per galaxy. 
\citep[e.g.,][]{katz91,nb91,katz92,sm94,sm95,nav97,wee98,sgv99,sn99,
eew00,koda00a,koda00b,aba03a,aba03b,meza03,sommer03,okamoto03,governato04,robertson04,okamoto05}.  
This mass range 
is almost the same or larger than the masses of the first collapsed objects ($\sim 10^6\Msun$) 
\citep[e.g.,][]{tegmark97,yoshida03}.
Therefore, these previous simulations could not address 
the hierarchical clustering processes 
from the first collapsed objects in the CDM universe.
It is also not clear how globular cluster-size objects 
are formed and evolved in forming galaxies.

In this paper, we study the evolution of clumps
in the halo of a host dwarf galaxy with high resolution $N$-body/SPH simulations.
Our cosmologically-motivated simulations are improved from previous simulations in 
terms of (1) particles masses, which are $\sim 10^3 \Msun$ for SPH particles and 
$\sim 10^4 \Msun$ for dark matter particles \citep[cf. $\sim 10^6 \Msun$ for SPH particles in][]{aba03a}, 
(2) inclusion of the low temperature radiative cooling below $10^4 K$, which allows formation of cold clouds.
Consequently, we can resolve structures larger than  $M_{\rm Jeans} \ge 2 N_{\rm nb} 
m_{\rm SPH}\ \sim 10^5 \Msun$ \citep{bat97} where $M_{\rm Jeans}$ is a Jeans mass,   
$N_{\rm nb} = 50$ is the number of neighbor particles used in the SPH calculation  
and $m_{\rm SPH}$ is the mass of an SPH particle.
Our effective mass resolution $M_{\rm Jeans} \sim 10^5 \Msun$ corresponds to the mass of baryonic 
substructure such as giant molecular clouds and GCs in galaxies.
In order to perform such high resolution simulation,  
we focus on the formation of a small galaxy (with total mass of $10^{10} \Msun$).

This paper is organized as follows:
We describe numerical method and models in \S \ref{sec:Model}.
In \S \ref{sec:CoolingEffects}, we demonstrate the effect of low-temperature cooling ($T < 10^4$ K) 
for the formation of clumps.
In \S \ref{sec:TidalEvolution},
we discuss the tidal evolution of clumps in the galactic halo.
Summary and discussion are given in \S \ref{sec:SummaryDiscussion}.

%
\section{Models and Numerical Methods}\label{sec:Model}
%

Our numerical technique is based on a standard $N$-body/SPH method \citep[e.g.,][]{hk89}, 
where self-gravity, radiative cooling and star formation from dense 
gas clouds are taken into account with a high spatial resolution ($\sim $ 100 pc).
In SPH simulations, the finer mass resolution enables us to 
resolve denser and colder structures.  
In fact, our particle mass ($M_{\rm SPH} \sim 10^3 \Msun$) allows us  
to follow the cold gas phase below $T = 10^4$ K.  
We thus consider gas cooling in the 
range between $10 ~{\rm K}$ and $10^8 ~{\rm K}$.
We review our models and numerical scheme in this section.

\subsection{Initial Conditions and Resolutions}\label{sec:Init}

We study galaxy formation from a top-hat density peak. 
Additional small-scale density perturbations are added using 
GRAFIC2 in the COSMICS package \citep{bertschinger01}.  
We use the following set of cosmological parameters:
$\Omega_0 = 1$, $\Omega_{\lambda} = 0.0$, $\Omega_b = 0.1$,
$h = 0.5$ (the unit is $100 {\rm km/s/Mpc}$), and $\sigma_8 = 0.63$.
The total mass of the sphere is $\sim 10^{10} \Msun$.

The initial redshift of the simulation, $z_{\rm ini}$, is $\sim 87$.
We impose an overdensity which corresponds to the collapse epoch $z_{\rm c} \sim 3$,
and the turnaround epoch $z_{\rm t}\sim 5.5$.
Since the limited volume cannot represent a realistic tidal field, 
matters in the region cannot get angular momentum from the external tidal fields. 
We thus initially add a rigid rotation for matters in the simulation sphere.
The spin parameter, $\lambda$, is $0.05$, 
which is a typical value derived from cosmological $N$-body simulations \citep[e.g.,][]{warren92}, 
while we cannot follow the angular momentum evolution correctly \citep{white84}. 

The total number of the particles is 
$N_{\rm DM} = N_{\rm SPH} = 1005600$.
The corresponding masses of particles are $m_{\rm DM} = 1.0 \times 10^4 \Msun$ for dark matter 
and $m_{\rm SPH} = 1.12 \times 10^3 \Msun$ for baryonic particles.
The gravitational softening lengths are comovingly evolved
from the beginning of the simulation to $z = 10$ \citep{governato04},
and they are fixed at 52 pc for baryonic particles 
and 108 pc for DM particles from $z = 10$ to $z = 0$.

Although the initial conditions we use are artificial and the adopted background cosmology is not a recently 
favored $\Lambda$-dominated universe, 
the background cosmology is not really important for the semi-cosmological 
simulations as far as the magnitudes of the additional small-scale perturbations are reasonable. 
Moreover at high-redshift ($z > 2$) where we study the evolution of small galaxies in this paper, 
the Einstein-de Sitter universe is still a good approximation.

The parameter set of this simulation is summarized 
in Table \ref{tab:Cosmology} (the cosmological parameters),
Table \ref{tab:Halo} (the parameters of the halo), 
and Table \ref{tab:Resolution} (the number of particles, 
mass resolutions and gravitational softening lengths of particles).

\begin{table}
\begin{center}
\caption{Cosmological parameters.}\label{tab:Cosmology}
\begin{tabular}{ccccc}
\hline
\hline
$\Omega_{\rm m}$ & $\Omega_{\rm \lambda}$ & $\Omega_{\rm b}$ & $h$ & $\sigma_8$ \\
\hline
$1.0$ & $0.0$ & $0.1$ & $0.5$ & $0.63$ \\
\hline
\end{tabular}

\caption{Parameters of the Halo.}\label{tab:Halo}
\begin{tabular}{ccccc}
\hline
\hline
$M_{\rm halo}$ & $z_{\rm ini}$ & $z_{\rm t}$ & $z_{\rm c}$ & $\lambda$ \\
\hline
$1.0 \times 10^{10} \Msun$ & $87$ & $5.5$ & $3.0$ & $0.05$ \\
\hline
\end{tabular}\\
Note. $H_0 = 100 h$ km/s/Mpc, where $H_0$ is a current Hubble constant.
$\sigma_8$ is the rms density fluctuations averaged
in spheres of 8 $h^{-1}$ Mpc radius at $z$ = 0.

\caption{Numerical Resolution}\label{tab:Resolution}
\begin{tabular}{cccc}
\hline
\hline
& particle number& particle mass & gravitational softening length\\
\hline
DM & 1005600 & $1.0 \times 10^4 \Msun$ & 108 pc \\
SPH  & 1005600 & $1.12 \times 10^3 \Msun$ & 52 pc \\
\hline
\end{tabular}
\end{center}
\end{table}

\subsection{Tree+GRAPE N-body/SPH code}
We accelerate gravity calculations by combining 
two fast gravitational solvers, i.e.,
a special purpose computer GRAvity PipE \citep[GRAPE;][]{sugimoto90}
and Tree algorithm \citep[e.g.,][]{appel85,bar86}.
The combination of GRAPE and Tree (hereafter Tree+GRAPE) 
is firstly proposed by \citet{makino91}  
and used in astronomical simulations \citep[e.g.,][]{fukushige01}.

Parameters used in our Tree+GRAPE algorithm are as follows:
the critical number of particles 
to share the same interaction list, $n_{\rm c}$, is $5000$ \citep[see][]{kawai00}.
The opening angle for Tree, $\theta$, is set to $0.5$ and we only use 
the monopole moment. 
The calculation speed of gravitational forces using our Tree+GRAPE code is
about dozens times faster than when we solely use Tree or GRAPE 
in tests on a system with a GRAPE and a host.

Gas dynamics is solved by the SPH method \citep[e.g.,][]{lucy77,GM77}, in which
shocks are handled via a standard artificial viscosity.
Viscosity parameters are $\alpha = 1.0$, $\beta = 2.0$ and $\eta = 0.1$ \citep{mon83}.
The number of neighboring particles for each SPH particle is set to $N_{\rm nb} = 50$.
We adopt a shear-reduced technique in the artificial viscosity \citep{bal95} .
We solve an energy equation of the ideal gas, $\gamma = 5/3$, 
with the radiative cooling and the inverse Compton cooling.
We assume that the gas has a primordial abundance of $X = 0.76$ and $Y = 0.24$, 
where $X$ and $Y$ are the mass fractions of hydrogen and helium.
The mean molecular weight is fixed at $\mu = 0.59$.

We apply the Tree+GRAPE method for neighbor search to reduce the computational cost.
We first generate a potential neighbor list using Tree,
and search neighboring particles from the list with GRAPE.
We adopt the reordering method to reduce 
the cost of data transfer between GRAPE and host computer \citep{sai03}.
The performance is an order of magnitude higher than Tree or GRAPE calculations.

\begin{figure}[h]
\begin{center}
\epsscale{0.9}
\plotone{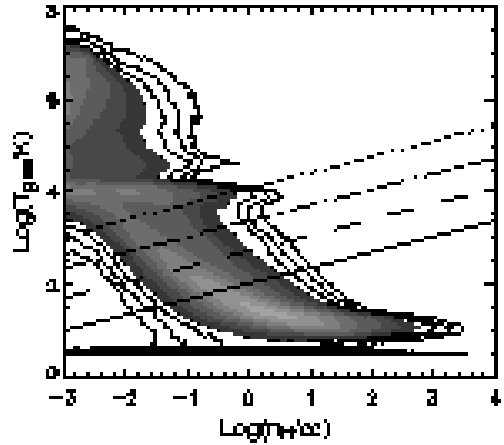}
\caption{
Phase diagram and limits of resolution.
The Jeans instability of the gas above the lines can be calculated properly in SPH simulations.
The solid, dashed, dashed-dotted, and dotted lines indicate
Jeans mass limits, $M_{\rm Jeans} = 2 N_{\rm nb} \times M_{\rm SPH}$, 
with the mass resolutions of a single SPH particle being
$M_{\rm {SPH}}=10^3\Msun$, $M_{\rm {SPH}}=10^4\Msun$, 
$M_{\rm {SPH}}=10^5\Msun$ and $M_{\rm {SPH}}=10^6\Msun$, respectively.
In the figure, we assume $N_{\rm nb} = 50$, $\mu = 0.59$, and $\gamma = 5/3$.
The contour map represents the volume-weighted phase diagram
from the numerical simulation of the multiphase ISM in a galaxy \citep{wad05}.
\label{fig:Jeans}
}
\end{center}
\end{figure}

\citet{bat97} and \citet{bate03} pointed out that, in the SPH simulations,
the Jeans instability cannot be calculated correctly for
masses less than $1.5-2 N_{\rm nb}  m_{\rm SPH}$,
where $m_{\rm SPH}$ and $N_{\rm nb}$ are the mass of an SPH particle and
the number of neighbor particles, respectively.
In our simulations, the thermal temperatures of gas particles are 
restricted to retain a Jeans mass larger than 
$2 N_{\rm nb} m_{\rm SPH}$, that is
\begin{eqnarray}
M_{\rm {Jeans}} &\sim& \rho \times \lambda_J^3, \\
&\sim& G^{-3/2} \rho^{-1/2} c_s^3 \ge 2 N_{\rm nb} m_{\rm {SPH}},
\end{eqnarray}
where $\lambda_J$, $G$, $c_s$, and $\rho$ are the Jeans length, 
gravitational constant, sound speed, 
and local density of gas, respectively.
Equivalently, the gas temperature should be greater than $T_{\rm min}(\rho)$, 
\begin{equation}
T_{\rm min}(\rho) = \frac{2 \mu m_{\rm p}}{3 k_{\rm B}} \frac{G \rho^{1/3} 
        (2 N_{\rm nb} m_{\rm SPH})^{2/3}}{\gamma (\gamma -1)} , \label{eq:Jeans:T}
\end{equation}
where $m_{\rm p}$ is the mass of a proton and $k_{\rm B}$ is the Boltzmann constant.
In Figure \ref{fig:Jeans}, 
we show critical lines of Jeans limits for several masses of SPH particles,
which is defined by Eq.(\ref{eq:Jeans:T}).
In our model, $m_{\rm SPH}$ is $1.12 \times 10^3 \Msun$, therefore
the minimum temperature of the gas with number density $n_{\rm H} \sim 100$ cm$^{-3}$ 
is $\sim 500$ K.

This high resolution in our simulations allows us to calculate the cold gas phase ($T \ll 10^4$ K), 
which is a potential site of star formation.
Previous numerical simulations of galaxy formation did not have 
sufficient resolution to calculate such a cold phase.
We adopt the cooling curve of \citet{spa97}, 
which is used in 2-D and 3-D simulations of the ISM on a galactic scale \citep[e.g.,][]{wad01,wad01b}.
We also include the inverse Compton cooling \citep{ikeuchi86}.
The allowable temperature for each gas particle with density $\rho$ 
is $T_{\rm min}(\rho) < T < 10^8$ K in this paper.
The radiative cooling below $10^4$ K plays an 
important role in the formation of clumps (see \S \ref{sec:CoolingEffects}). 

\subsection{Star Formation}
Our star formation algorithm is similar to the one by \citet{katz92}.
We adopt a {\it conversion} type star formation.
A gas particle which satisfies all the following conditions
is eligible to be converted to a star particle:
The gas particle 
(1) has high density ($n_{\rm H} > 0.1 ~{\rm cm^{-3}}$),
(2) lies in an overdense region ($\rho_{\rm gas} > 200 \times \rho_{\rm BG}(z)$), 
    where $\rho_{\rm BG}(z)$ is the cosmic background density at $z$, 
(3) has low temperature (T$<30,000$K), 
and (4) is in a collapsing region ($\nabla \cdot v < 0$).

The star formation rate of a gas particle is given by the Schmidt law, 
\begin{equation}
\frac{d \rho_{*}}{dt} = c_{*} \frac{\rho_{\rm gas}}{t_{\rm dyn}},
\end{equation}
where 
$\rho_{*}$ is the density of newly-born stars,
the local star formation efficiency, $c_{*}$, is $1/30$ \citep[e.g.,][]{aba03a}
and the dynamical time is $t_{\rm dyn} = 1/\sqrt{4 \pi G \rho_{\rm gas}}$.
Then, the probability of turning the gas particle into a star particle during a time step $\delta t$ 
is given by 
\begin{equation}
p = \Bigl [ 1 - \exp \Bigl( - c_{\rm *} \frac{\delta t}{t_{\rm dyn}}  \Bigl)  \Bigl ].
\end{equation}
We do not include the mechanical and radiative feedback from star formation, SNe 
and succeeding metal enrichment to the gas for simplicity.

%
\section{Effects of Low Temperature Cooling}\label{sec:CoolingEffects}
%

In order to demonstrate the importance of the low temperature cooling ($T < 10^4$ K),
we run two simulations with and without the cooling below $T < 10^4$ K from the same initial condition.
We call the former simulation `model A' and the latter, `model B'.

\begin{figure*}[t]
\begin{center}
\epsscale{1.0}
\plotone{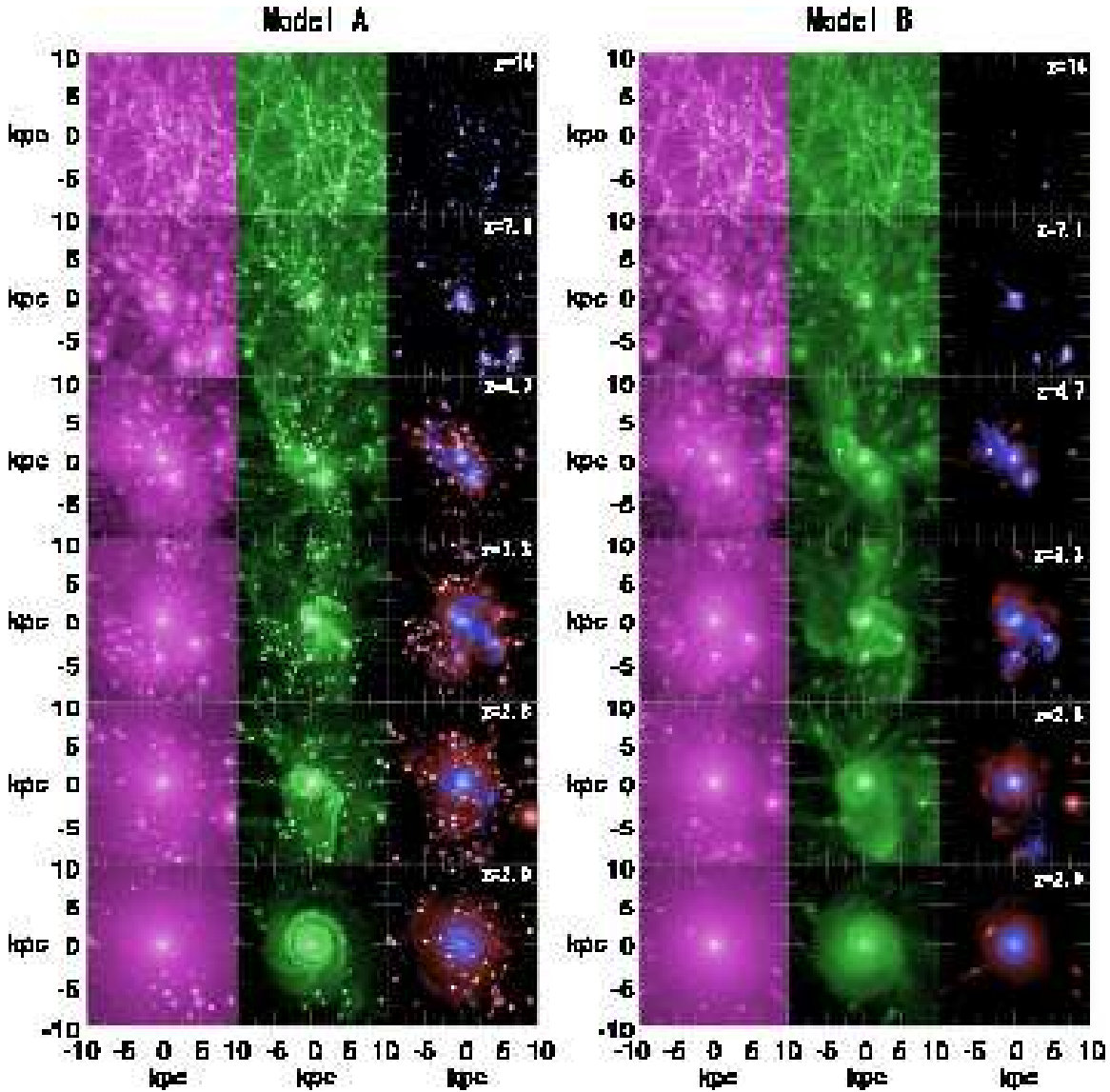}
\caption{
Effects of low temperature cooling ($T < 10^4$ K)
on the clump formation in a galaxy.
In model A, the gas can cool below $T < 10^4$ K,
whereas the cooling above $T = 10^4$ K is assumed in model B.
Left, middle, and right panels in each model show the distribution of 
dark matter, gas, and star particles, respectively.
Blue and red in each right panel indicate 
young (age $\le$ 0.1Gyr) and old star particles, respectively.
Brightness of each particle represents the local density of each particle.
\label{fig:SnapShot}
}
\end{center}
\end{figure*}

The snapshots of model A and B are shown in Figure 
\ref{fig:SnapShot}.
The figure shows dark matter (left panels),
gas (central panels) and stars (right panels) at six different epochs from $z = 14$ to $2$.
Clearly, the clumpiness of the gas is very different between the two simulations.
Model A has many clumps throughout the formation of the galaxy,
whereas the distribution of gas in model B is smooth,
because the gas cannot cool below $10^4$ K.  
In other words, small scale structures cannot be evolved due to high sound speed. 
The number of clumps in model B is an order of magnitude fewer than that in model A
(see \S \ref{sec:DefClumps}). 
Table \ref{tab:NumberMass} shows the numbers and mass fractions of clumps.
The total masses contained in clumps 
do not change significantly between $z = 4.7$ and $3.3$ in both models.
Since massive clumps quickly sink 
and merge into the central galaxy between $z = 3.3$ and $2.0$,  
the mass fraction decreases to $z = 2$. 

Model A shows a variety of fine structures at $z = 2$,
such as spirals, clumps, tidal tails,
while model B does not show such features.
These structures consist largely of the low temperature gas.
Gas cooling below $10^4$ K is important for investigating the evolution of clumps,
and consequently the evolution of host galaxies.
In the following sections, we discuss evolution of the clumps based on the results of model A.

\begin{table}
\begin{center}
\caption{Number and Mass Fraction of Clumps.}\label{tab:NumberMass}
\begin{tabular}{ccccc}
\hline
\hline
 & \multicolumn{2}{c}{Model A} &  \multicolumn{2}{c}{Model B} \\
 & Number & Fraction & Number & Fraction \\
\hline
$z=4.7$ & $54$ & $5.5$\% & $14$ & $2.5$\% \\
$z=3.3$ & $87$ & $6.6$\% & $22$ & $4.2$\% \\
$z=2.0$ & $62$ & $2.6$\% & $7$ & $0.12$\% \\
\hline
\end{tabular}
\end{center}

`Number' and `Fraction' indicate the number of clumps within the virial radius of the galaxy
and mass ratio of total mass of clumps to the mass of the galaxy, respectively.
\end{table}

%
\section{Tidal Stripping of Dark Halos around Clumps}\label{sec:TidalEvolution}
%

In this section, we investigate the evolution of clumps 
in the galactic tidal field of model A.
We firstly define clumps and construct the evolutional history of each clump.
we then quantitatively investigate the effect of tidal stripping 
for the dark halos around the clumps.
We compare stripping effects on clumps in three different epochs, 
i.e., before the collapse epoch ($z = 4.7$), during the collapse epoch ($z = 3.3$)
and after the collapse epoch ($z = 2.0$) to see the evolution.

\subsection{Identification of Clumps}\label{sec:DefClumps}

In order to identify clumps,
we use SKID \citep{governato97} as a clump finder, which is
based on a local density maxima search.
This algorithm groups particles by moving them along the density 
gradient to the local density peaks. The density field and density gradient 
are defined everywhere by smoothing each particle using the SPH-like method 
with 64 neighboring particles.  
At each density peak, the localized particles are grouped into an initial particles list of the clump.
Then, gravitationally-unbound particles are removed from the initial particles list of the clump.
If the number of the bound particles is more than a threshold number, $n_{\rm th}=100$, we identify 
this group of particles as a `clump'.
We define the positions of each clump as its center of mass and the size of each clump, $r_{\rm sub}$, as 
the distance of the particle furthest from the center.

\begin{figure}[h]
\begin{center}
\epsscale{1.0}
\plotone{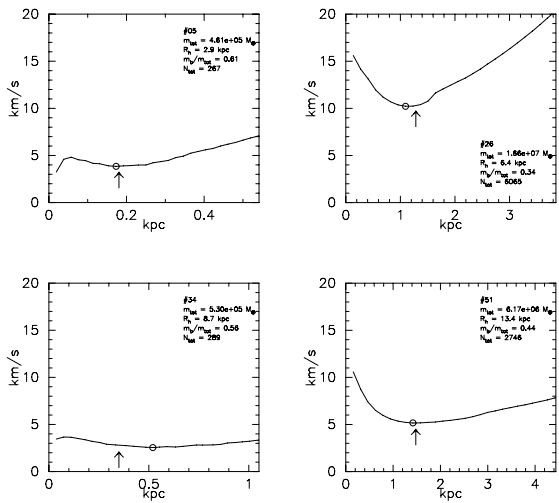}
\caption{
Circular velocity profiles of randomly selected four clumps at $z = 2.0$.
The circular velocities are estimated by $\sqrt{G M(<r)/r}$, 
where $r$ is the distance from the center of the clumps and $M(<r)$ is the mass within $r$, respectively.
We plot three times of $r_{\rm sub}$ for $r$.
The upward-pointing arrows indicate the positions of $r_{\rm sub}$s and 
the open circles represent the the minimum points of the circular profiles, respectively.
\label{fig:circulars}
}
\end{center}
\end{figure}

In Figure \ref{fig:circulars},
we plot circular velocity profiles of randomly selected four clumps.
The circular velocities are defined as $\sqrt{G M(<r)/r}$, 
where $r$ is the distance from the center of the clumps
and $M(<r)$ is the mass within $r$, respectively.
The upward-pointing arrow in each panel indicates 
the clump size which is derived from SKID, namely $r_{\rm sub}$.
The $r_{\rm sub}$s are comparable with the minimum points of the circular velocity profiles (the open circles),
i.e., $r_{\rm sub} \simeq r(v_{\rm c} = {\rm min})$.
Thereby, our clump finding method successfully finds the size of the clumps.

\subsection{Distribution of Dark Matter and Baryons in Clumps}

\begin{figure}[h]
\begin{center}
\epsscale{1.0}
\plotone{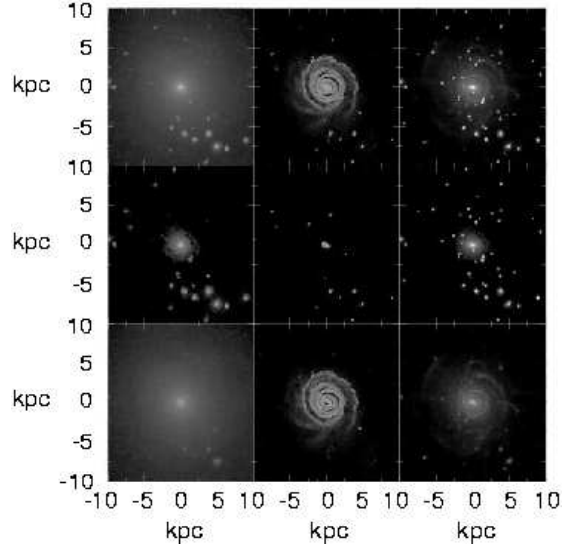}
\caption{
Demonstration of our clump finding algorithm. 
Snapshots of dark matter (left column), gas (center column) 
and star (right column) particles at $z = 2$ are shown.
From top to bottom, we show particles with three different criteria:
all particles of each component (top row),
the particles in the clumps which are detected by SKID (middle row)
and the residual particles (bottom row).
To highlight structures in the halo, we plot the particles 
with the weighted brightness proportional to the local densities of the particles.
\label{fig:skid}
}
\end{center}
\end{figure}

Figure \ref{fig:skid} shows the particle distributions at $z=2$.
The particles in the clumps are plotted in the middle row.
The three panels in the bottom row are residual particles from the
original distribution shown in the top three panels.
The positions of the three components, dark matter, gas, and starts,
coincide with each other. 
However, the size of these components differs,
in a sense that gas and stars are centrally concentrated 
and the dark matter halos are extended. 
This is a natural consequence of dissipational nature of the gas. 
It is remarkable that some clumps do not have dark matter halos at $z = 2$
(see the first quadrant in the middle row of Figure \ref{fig:skid}).

\subsection{Evolution of Clumps}\label{sec:orbit}
In order to trace the evolution of clumps,
we construct evolution histories of clumps.
Time evolution of each  clump is tracked in the following way:
At first, we prepare two clump catalogs (catalog $i$ and $i+1$)
at two adjacent time snapshots.
If a clump in $i$ has more than a half of particles in a clump in $i+1$, 
it is regarded as the progenitor of the clump in catalog $i+1$.
We trace clumps in our result back to $z = 15$, using 
235 snapshots (time interval of $10^7$ yr), and we obtain the evolutional tracks of 
all clumps found at $z = 2.0$.

\begin{figure}[h]
\begin{center}
\epsscale{1.0}
\plotone{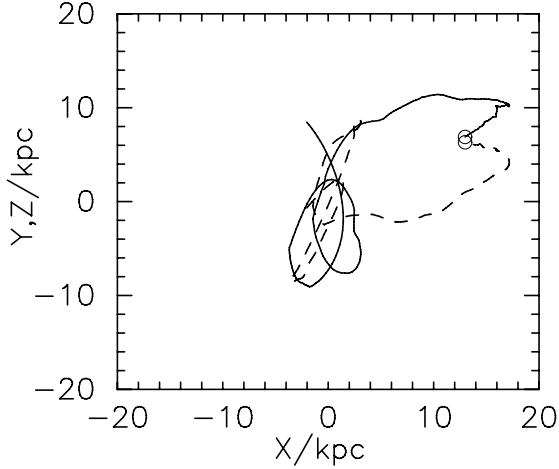}
\caption{
Projected orbits of the reference  clump.
Solid and dashed lines indicate the trajectories on $x-y$ and $x-z$ planes, respectively.
Origin of coordinates accord with the center of the galactic halo.
Two circles show the starting points of each trajectory.
\label{fig:xyzevolution}
}
\end{center}
\end{figure}

\begin{figure}[h]
\begin{center}
\epsscale{1.0}
\plotone{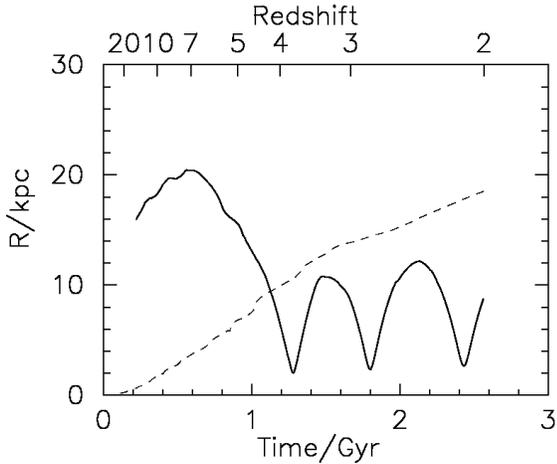}
\caption{
Distance of the reference clump from the galactic center, $R_{\rm h}$, 
and the virial radius of the host galaxy, $R_{\rm vir}$, against time or redshift.
Solid and dashed lines indicate $R_{\rm h}$ and $R_{\rm vir}$, respectively.
\label{fig:rhevolution}
}
\end{center}
\end{figure}

Figure \ref{fig:xyzevolution} shows the  trajectories of 
a clump on $x-y$ and $x-z$ planes.
This is a typical case of a clump being captured by the host halo.
Figure \ref{fig:rhevolution} shows 
the distance from the center of mass of the clump to the galactic center 
$R_{\rm h}$ (solid line) and the virial radius of the host halo $R_{\rm vir}$ (dashed line)
against time and redshift.
The clump gets into the virial radius of the host galaxy at $z \sim 4.5$.
The pericenter-to-apocenter ratio of the elliptical orbit is $\sim 0.2$, 
which is consistent to the mean value of observed galactic globular clusters 
\citep[e.g.,][;the mean value is $\sim 0.3$]{Dinescu99}
and the values in clusters in CDM simulations \citep[e.g.,][]{ghigna98,okamoto99,gill04}.
In our analysis shown below, we refer to this clump as the reference case.
{\footnote 
{The clump labeled \#34 in Figure \ref{fig:circulars} (bottom-left) is the reference clump.}
}

\subsection{Size of Clumps}
Since clumps in the galactic halo are subjected to a tidal force of the host galaxy,
their sizes, $r_{\rm sub}$, should correlate with a tidal radius which is determined by the tidal force.
Assuming isotherm density profiles [$\rho(r) \propto r^{-2}$] for the host halo and clumps, 
the tidal radius for a clump at $R_{\rm h}$ is defined by
\begin{equation}
r_{\rm tidal}(R_{\rm h}) \equiv R_{\rm h} 
\sqrt{\frac{R_{\rm h} \ m_{\rm tot}}{r_{\rm sub} \ M_{\rm tot}(R_{\rm h})}} \label{eq:TidalRadius},
\end{equation}
where $m_{\rm tot}$ and $M_{\rm tot}(r<R_{\rm h})$ are 
the total mass of the clump and the total mass of the host galaxy within $R_{\rm h}$, 
respectively \citep{ghigna98, okamoto99}.
This equation is derived from the balance between 
the tidal force due to the host halo and the self-gravity of the clump.

\begin{figure}[h]
\begin{center}
\epsscale{1.0}
\plotone{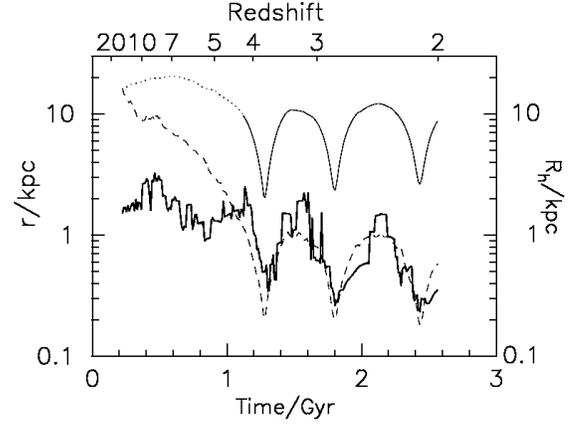}
\caption{
Time evolution of the size of the reference clump ($r_{\rm sub}$)
and tidal radius ($r_{\rm tidal}$).
Thick-solid and dashed lines indicate $r_{\rm sub}$ and $r_{\rm tidal}$, respectively.
Distance of the clump from the galactic center is shown by the thin-solid line (see Fig. \ref{fig:rhevolution}).
The dotted line indicates $R_{\rm h}$ before capture of the host galactic halo.
\label{fig:rrrlevolution}
}
\end{center}
\end{figure}

Evolution of $r_{\rm sub}$, $r_{\rm tidal}$ of the reference clump, 
and $R_{\rm h}$ is shown in Figure \ref{fig:rrrlevolution}.
The clump is captured by the host galactic halo at $z \sim 4.5$.
The size of the clump is independent of 
the tidal radius before $z \sim 4.5$.
However, it follows $r_{\rm tidal}$ after $z \sim 4.5$.
This suggests that once a clump is captured by the host halo, 
its radius instantaneously shrinks to the tidal radius.

\begin{figure}[h]
\begin{center}
\epsscale{1.0}
\plotone{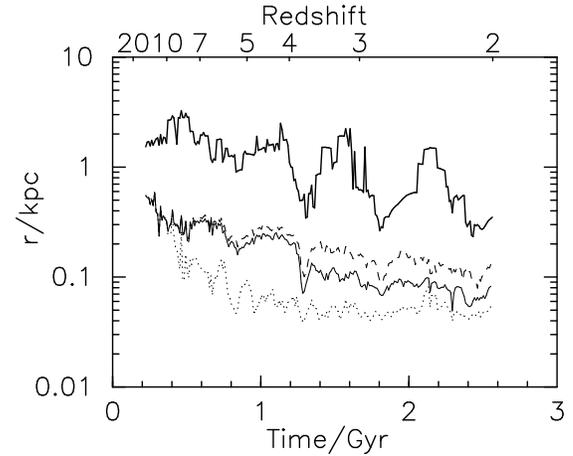}
\caption{
Time evolution of the half mass radii of the total, baryon, and dark matter components
of the reference clump.
Thin-solid, dotted and dashed lines indicate 
$r_{\rm hlf,tot}$, $r_{\rm hlf,b}$, and $r_{\rm hlf,dm}$, respectively.
Thick-solid line indicates $r_{\rm sub}$, which is already shown in Figure \ref{fig:rrrlevolution}.
\label{fig:rhlf}
}
\end{center}
\end{figure}

In Figure \ref{fig:rhlf} we display the half mass radii for total ($r_{\rm hlf,tot}$), 
baryon ($r_{\rm hlf,b}$), and dark matter ($r_{\rm hlf,dm}$) components  
as representative sizes of the distributions of matters within the clump.
From the plot, we find that 
(1) $r_{\rm hlf,b}$ is always smaller than $r_{\rm hlf,dm}$
, and (2) $r_{\rm hlf,tot}$ comes close to $r_{\rm hlf,b}$ with time 
while $r_{\rm hlf,tot}$ is almost similar to $r_{\rm hlf,dm}$ before the first pericenter passage ($z\gtrsim 3.7$).
Due to the radiative cooling, the gas shrinks 
towards the center of the clump immediately and forms a core in there. 
This is the reason of the first point.
The second point, namely $r_{\rm hlf,tot}$ shifts from $r_{\rm hlf,dm}$ to $r_{\rm hlf,b}$,
is explained as the effect of the tidal stripping.

\begin{figure*}[t]
\begin{center}
\epsscale{1.0}
\plotone{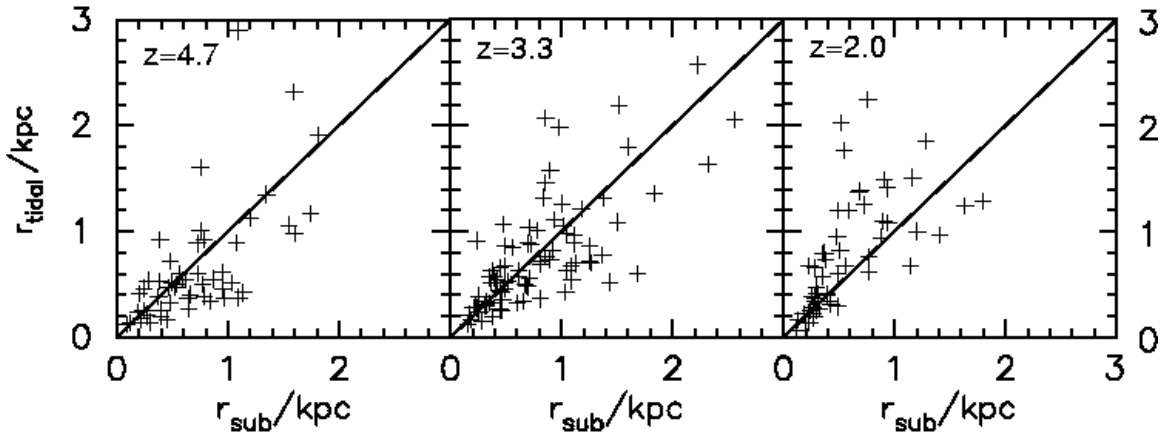}
\caption{$r_{\rm tidal}$ against $r_{\rm sub}$.
The black line indicates that both quantities coincide.
\label{fig:tidal}
}
\end{center}
\end{figure*}

In Figure \ref{fig:tidal}
we plot $r_{\rm tidal}$ of clumps in the galactic halo 
against $r_{\rm sub}$ at $z = 4.7, 3.3$ and $2.0$.
If the radii of clumps are determined by the tidal field, 
we should find a correlation (solid line) in Figure \ref{fig:tidal}.
These figures clearly indicate that the sizes of clumps are
determined by the local tidal field instantaneously,
as we have already shown in Figure \ref{fig:rrrlevolution},
in spite of the tidal stripping being more efficient at the pericenter.
This is because the matters off from clumps spill 
their local tidal radii after passing away from the pericenter.
Indeed, some of the clumps show clear evidence of elongation as the tidal tails
(see the distribution of old stars in model A in figure \ref{fig:SnapShot})

\subsection{Tidal Truncation of Dark Halos and Formation of Pure Baryonic Clumps}
\label{subsec:mbmt}

Since the baryons have more concentrated distribution than the dark matter within a clump,  
tidal stripping should affect them differently. 
We will show this effect more quantitatively below.

\begin{figure}[b]
\begin{center}
\epsscale{1.0}
\plotone{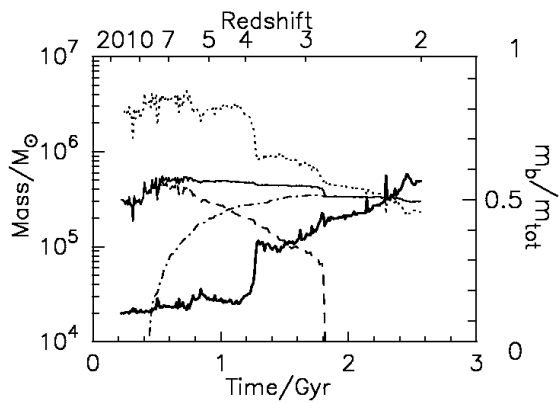}
\caption{
The evolution of masses of the reference clump
and $m_{\rm b}/m_{\rm tot}$ of the clump.
Thin-solid, dashed, dot-dashed, and dotted lines indicate 
the masses of baryon, gas, stars and dark matter in the clump, respectively.
Thick-solid line shows the evolution of $m_{\rm b}/m_{\rm tot}$ of the clump.
\label{fig:massmbmtevolution}
}
\end{center}
\end{figure}

Figure \ref{fig:massmbmtevolution} displays the mass evolution of 
three components (gas, star and the dark matter) of the reference clump.
The mass of dark matter decreases with time. 
Mass loss occurs mainly at each pericenter passage once a clump is captured.
The rapid mass loss, for example at around $t \sim 1.3$ Gyr ($z \sim 3.7$), 
corresponds to the time when the clump is passing the pericenter.  
The gas/stellar mass decreases/increases steadily owing to the star formation.
However, the total mass of baryons (gas + star) is almost constant.  
This is because that the baryonic component tends to reside
inside the tidal radius of the clump and is therefore less vulnerable
to tidal stripping.
Eventually, the total mass of the clump becomes one-tenth of its original mass.
As shown in Figure \ref{fig:massmbmtevolution}, the baryon mass to total mass ratio 
in the reference clump increases with time, up to 0.5.

\begin{figure}[b]
\begin{center}
\epsscale{1.0}
\plotone{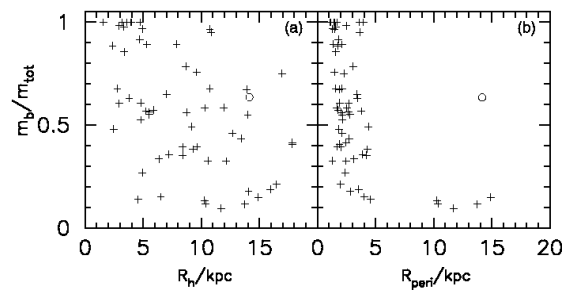}
\caption{ 
Distribution of $m_{\rm b}/m_{\rm tot}$ in the galactic halo at $z = 2.0$.
The left panel shows $m_{\rm b}/m_{\rm tot}$ against $R_{\rm h}$.
On the other hand, the right panel shows
$m_{\rm b}/m_{\rm tot}$ against the pericentric distance, $R_{\rm peri}$.
Crosses in the panels represent each clump.
The open circle in the right panel is the clump which is preprocessed by the other halos.
\label{fig:MbMt}
}
\end{center}
\end{figure}

The baryon mass fractions in clumps in the galactic halo are plotted against 
$R_{\rm h}$ in Figure \ref{fig:MbMt} (a) for a snapshot at $z = 2$.
Many clumps show signs of tidal stripping 
($m_{\rm b}/m_{\rm tot} > 0.1$).
Since tidal force from the host galaxy is effective for clumps with smaller $R_{\rm h}$,
$m_{\rm b}/m_{\rm tot}$ should be larger for smaller $R_{\rm h}$.
However, there is no clear trend between $R_{\rm h}$ and $m_{\rm b}/m_{\rm tot}$ in Figure \ref{fig:MbMt} (a).
Suppose that the tidal stripping is the most effectively occurs following each pericenter passage, 
we expect that $m_{\rm b}/m_{\rm tot}$ is proportional to the inverse of $R_{\rm peri}$.
In fact, Figure \ref{fig:MbMt} (b) shows this trend.
{\footnote {We adopt the current distance from the center of the halo, $R_{\rm h}$, instead  
of $R_{\rm peri}$, if a clump is in the first falling phase.}}
Note also that no clump is found within $R_{\rm peri} < 1$ kpc,
since the dynamical friction time scales of clumps within 1 kpc ($< 10^8$ yr) 
are smaller than the age of the system ($\sim 10^9$ yr) and 
clumps with small pericenters ($<1$ kpc) fall into the galactic center and are tidally destroyed/disappeared.

\begin{figure}[h]
\begin{center}
\epsscale{1.0}
\plotone{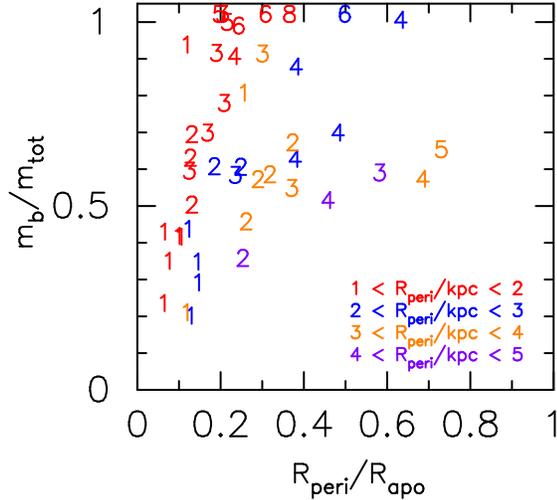}
\caption{ 
Distribution of $m_{\rm b}/m_{\rm tot}$ against $R_{\rm peri}/R_{\rm apo}$
in the galactic halo at $z = 2.0$.
Red, blue, orange, and purple numbers
indicate different ranges of the pericentric distance from
$1 < R_{\rm peri}/{\rm kpc} < 2$ to $4 < R_{\rm peri}/{\rm kpc} < 5$
and the numbers indicate how many times each clump passes the pericenter.
We obtain pericenters and apocenters from the most recent period of clumps.
Note that we does not consider clumps in this plot, which have never passed pericenters.
and larger pericentric distances, $R_{\rm peri} > 5$ kpc.
\label{fig:RperiRapoMbMt}
}
\end{center}
\end{figure}

Another possible effect to determine $m_{\rm b}/m_{\rm tot}$ is the eccentricity of each clump's orbit.
Even if $R_{\rm peri}$ is the same, clumps with high eccentricities spend 
most of their time at apocenter ($R_{\rm apo}$), 
as a result they are less severely stripped than those with smaller eccentricities.   
In Figure \ref{fig:RperiRapoMbMt}, we plot $m_{\rm b}/m_{\rm tot}$ 
as a function of $R_{\rm peri}/R_{\rm apo}$ at $z = 2$.
As expected, there is a weak positive correlation 
between $m_{\rm b}/m_{\rm tot}$ and $R_{\rm peri}/R_{\rm apo}$.
In addition to that, the plot also suggests that the baryon fraction is determined by 
(1) how many times the clumps pass the peri-center, represented by the numbers, 
and (2) $R_{\rm peri}$, represented by the color sequences.

\begin{figure}[h]
\begin{center}
\epsscale{1.0}
\plotone{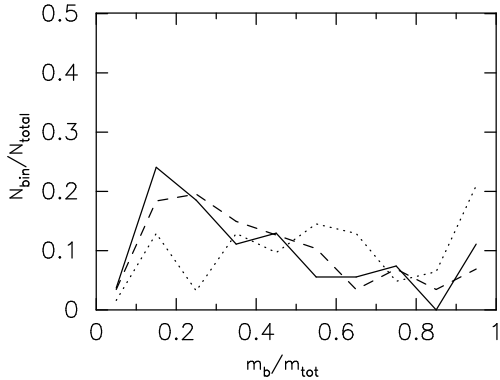}
\caption{
Fraction of clumps with $m_{\rm b}/m_{\rm tot}$
at three redshifts, i.e., 
$z = 4.7$(solid line), $z = 3.3$(dashed line), and $z = 2.0$(dotted line).
\label{fig:NormalizedMbMt}
}
\end{center}
\end{figure}

Figure \ref{fig:NormalizedMbMt} shows distribution of $m_{\rm b}/m_{\rm tot}$ 
in clumps at $z = 4.7, 3.3$ and $2.0$.
The vertical axis shows the number of clumps in each bin (0.1)
normalized by the total number of clumps at each epoch.
The total number of clumps is shown in Table \ref{tab:NumberMass}.
The fractions of baryon dominated systems ($m_{\rm b}/m_{\rm tot} > 0.5$) 
increases with time as  30 \%($z = 4.7$), 32 \%($z = 3.3$), and 60 \%(z = 2.0).
This is a natural consequence of the effective stripping of the dark matter halos 
from clumps by the tidal interaction with the host galaxy 
(see Figure \ref{fig:massmbmtevolution}).

%
\section{Summary and Discussion}\label{sec:SummaryDiscussion}
%

In order to study the co-evolution of galaxy and its substructure such as GC,
we run high-resolution $N$-body/SPH simulations of galaxy formation.
The numerical resolution of our simulation is high enough ($\sim 50$ pc and $\sim 10^3 \Msun$) 
to resolve the Jeans instability of small clumps whose mass is larger than $10^5 \Msun$. 
With this high resolution, radiative cooling below $10^4$ K 
can be taken into account and this low temperature cooling strongly affects evolution of these 
systems.
What we found on the evolution of clumps is summarized below:

\begin{itemize}

\item Gravitationally-bound, small clumps with core-halo structure are first formed from CDM perturbations.
    After captured by the galactic halo,
    these clumps lose their dark matter halo due to the tidal field of the host galaxy,
    and result in baryon-dominated objects, such as GCs.
    By $z = 2.0$, almost half of the clumps (57 \%) 
    becomes baryon-dominated systems ($m_{\rm b}/m_{\rm tot} > 0.5$).

\item The simulation without radiative cooling below $T = 10^4$ K (the model B) shows 
    an order of magnitude fewer clumps.
    This result implies that treatment of the low temperature cooling as well as 
    heating is extremely important to investigate the formation of low mass objects, such as GCs.

\end{itemize}

Clumps studied here have 
a mass range similar to that of globular clusters (GCs) or dwarf galaxies.
Two formation scenarios for GC formation has been proposed in literature.
One is that GCs are formed in DM minihalos 
before they infall into galactic halos 
\cite[e.g.,][]{Peebles84,bromm02,weilpudritz01,beasley03,Mashchenko2005a,Mashchenko2005b}.
The other is that GCs are formed from baryonic processes, 
such as hydrodynamical shocks, and thermal instability, in galactic halos
\citep[e.g.,][]{schweizer87,FR85,whitmore95,kravtsov05}.
Clumps around a host galaxy in our simulation 
(see the last panel of Figure \ref{fig:SnapShot})
originate in hierarchically formed DM minihalos, and therefore  
our numerical results support the former scenario.

The absence of DM halos around observed GCs \citep{pryor89,moore96}
could be the result of tidal stripping by interactions with the host galaxy.
\citet{Mashchenko2005b} studied the tidal evolution of primordial GCs 
with dark matter halos in a static potential of a host galaxy.
In their work, {\it dark matter-less} GCs form from proto-GCs with dark matter halos
via tidal stripping.
Our results also suggest that GCs can be formed naturally 
in the context of the cold dark matter universe.

In taking some heating mechanisms such as 
SN explosions and/or ultraviolet background heating and/or reionization 
\citep[e.g.,][]{Dekel86,Efstathiou92,sgv99,tc01,susa04a,susa04b}
into consideration,
the suppression and/or destruction of small mass clumps, such as GCs and dwarfs, should occur.
We plan a new series of simulations including feedback from stars,
without sacrificing the spacial and mass resolutions.

\acknowledgments
We thank an anonymous referee for valuable comments that considerably improved the manuscript.
This paper is based on a part of the author's (TRS) dissertation 
submitted to Hokkaido University, 
in partial fulfillment of the requirement for the doctorate.
J.K. and T.O. are financially supported by the Japan Society
for the Promotion of Science for Young Scientists.
Numerical computations were carried out on GRAPE clusters (MUV) (project ID:g03a12/g04a07)
at the Astronomical Data Analysis Center of
the National Astronomical Observatory, Japan.

\end{document}